\documentclass[a4paper,twoside,reqno]{amsart}

\usepackage[pagewise]{lineno}
\usepackage[utf8]{inputenc}

\usepackage{authblk}
\usepackage{hyperref}
\hypersetup{urlcolor=blue, citecolor=red}
\usepackage{amsmath,amsthm,amssymb,xcolor,bbm,mathrsfs,graphicx, float, orcidlink}
\usepackage{enumitem}
\usepackage[english]{babel}
\usepackage{fancyhdr}

\title{Asymptotic Stability of Conservative Convex-Combination Dynamics on Multilayer Graphs}
\author{Hsin-Lun Li}

\date{}
\email{hsinlunl@math.nsysu.edu.tw}

\newtheorem{theorem}{Theorem}[section]
\newtheorem{lemma}[theorem]{Lemma}
\newtheorem{corollary}[theorem]{Corollary}

\theoremstyle{definition}

\begin{document}

\allowdisplaybreaks

\thispagestyle{firstpage}
\maketitle
\begin{center}
    Hsin-Lun Li \orcidlink{0000-0003-1497-1599}
   \centerline{$^1$Department of Applied Mathematics, National Sun Yat-sen University, Kaohsiung 804, Taiwan}
\end{center}
\medskip

\begin{abstract}
We study discrete-time consensus dynamics on multilayer networks in which each layer evolves via a time-varying doubly stochastic interaction matrix, and inter-layer coupling is introduced through two mechanisms: (i) distribute-then-average and (ii) average-then-distribute. These define conservative redistribution processes that preserve total mass across all layers and can be viewed as stochastic averaging driven by products of time-inhomogeneous stochastic matrices with structured coupling.

For both mechanisms, we construct quadratic Lyapunov functionals that form nonnegative supermartingales, yielding almost sure convergence. The analysis combines martingale arguments with dissipation identities and connectivity properties of induced interaction graphs. Under recurrent connectivity conditions on subgraphs of the time-varying interaction structure, we prove asymptotic consensus to the global average determined by the initial total mass.

This provides a unified framework for multilayer averaging dynamics, extending classical consensus results for products of stochastic matrices to settings with explicit inter-layer coupling. As corollaries, we specialize the general framework to the multilayer garbage disposal dynamics, thereby establishing convergence guarantees under natural connectivity conditions on the underlying graphs.
\end{abstract}

\section{Introduction}

Multilayer networks provide a general framework for representing systems in which interactions occur simultaneously across multiple relational channels, and have been widely studied in the literature on complex systems \cite{boccaletti2014structure,kivela2014multilayer,dickison2016multilayer}. In such systems, dynamical processes are often driven by local averaging or redistribution mechanisms that can be represented as products of stochastic matrices, where convergence is governed by mass conservation and repeated mixing across time-varying interaction graphs. Classical consensus models of this type originate from the DeGroot framework \cite{degroot1974reaching} and have been extensively developed in networked multi-agent systems under switching or time-dependent topologies \cite{jadbabaie2003coordination,olfati2004consensus,moreau2005stability,olfati2007consensus}. More generally, the asymptotic behavior of such nonhomogeneous averaging processes is fundamentally rooted in the theory of non-negative matrices and stochastic matrix products, where ergodicity, weak contraction, and connectivity conditions play a central role \cite{seneta2006non,hajnal1958weak,wolfowitz1963products,touri2012product}. In particular, distributed averaging systems can be interpreted as time-varying Markov chains, and their convergence is characterized by classical ergodicity and infinite-flow conditions developed in \cite{touri2010ergodicity,touri2012product}.

Within social and behavioral dynamics, related averaging mechanisms appear in models of opinion formation and interaction-driven evolution, including classical bounded or structured influence processes \cite{friedkin1990social,castellano2009statistical}. In parallel, recent work has extended such dynamics to more complex interaction structures, including mixed and generalized Hegselmann--Krause type models and their convergence properties on infinite or structured graphs \cite{lanchier2022consensus,li2022mHK,li2023mHK2,li2024mixed,li2025leader}. Motivated by redistribution problems involving undesirable resources (“bads”), strategic formulations such as garbage allocation and exchange games have also been studied in economic and game-theoretic settings \cite{hirai2006coalition}, and more recently extended to multilayer environments \cite{li2025multilayer}.

The present paper introduces a conservative convex-combination framework for multilayer redistribution dynamics in which each layer evolves via a doubly stochastic interaction matrix, and inter-layer coupling is achieved through either distribution-then-cross-layer-averaging or cross-layer-averaging-then-distribution mechanisms. The resulting dynamics preserve total mass across layers and can be interpreted as a multilayer extension of stochastic averaging processes with structured coupling. The main results establish asymptotic consensus to the global average under recurrent connectivity conditions, thereby providing a theoretical foundation for the multilayer garbage disposal game introduced in \cite{li2025multilayer}.

\section{Multilayer Conservative Model}
We begin by introducing the notation and graph-theoretic constructions used in the multilayer dynamics. Let $[n]=\{1,\ldots,n\}$. For a matrix $A\in\mathbb{R}^{n\times n}$, denote by $A_{ij}$ its $(i,j)$-th entry. A matrix $A$ is called \emph{row stochastic} if $\sum_{j\in[n]} A_{ij}=1$ for all $i\in[n]$, \emph{column stochastic} if $\sum_{i\in[n]} A_{ij}=1$ for all $j\in[n]$, and \emph{doubly stochastic} if it is both row and column stochastic.

For a graph $G$, let $V(G)$ and $E(G)$ denote its vertex and edge sets, respectively. Given a graph $G$ with vertex set $V(G)=[n]$, the \emph{in-neighborhood} of vertex $i$ is defined by
\[
N_{G,i}=\{j\in[n]:(j,i)\in E(G)\}.
\]
For a matrix $A\in\mathbb{R}^{n\times n}$, the associated directed graph $G(A)$ has vertex set $[n]$ and edge set
\[
E(G(A))=\{(j,i):A_{ij}\neq 0\}.
\]
The \emph{order} of a graph is its number of vertices. For a graph $G$ of order $n$, we define an undirected graph $M(G)$ with vertex set $[n]$ and edge set
\[
E(M(G))=\{(i,j):N_{G,i}\cap N_{G,j}\neq\emptyset\}.
\]
Thus, two vertices are adjacent in $M(G)$ if and only if they share at least one common in-neighbor in $G$.

We consider a multilayer network with $n$ agents and $m$ layers. Each agent $i$ holds a state $x_i^{(\ell)}(t) \in \mathbb{R}$ in layer $\ell$ at discrete time $t \ge 0$, representing a generic scalar quantity (e.g., opinion, belief, resource level, or load such as garbage amount). The redistribution within each layer is described by a matrix $A^{(\ell)}(t)\in \mathbb{R}^{n\times n}$, whose $(i,j)$-th entry $A_{ij}^{(\ell)}(t)$ represents the proportion of agent $j$'s state redistributed to agent $i$ at time $t$. We assume that $A^{(\ell)}(t)$ is doubly stochastic for all $\ell\in [m]$ and $t\ge 0.$ Let 
\[
\bar{A}(t)=\frac1m\sum_{\ell\in[m]}A^{(\ell)}(t)
\]
be the average of $A^{(\ell)}(t)$ over $\ell\in[m]$.  

We study two natural mechanisms for updating agent states across layers.  

\paragraph{Mechanism I: Distribute-then-Average.}  
In each layer $\ell$, agents redistribute their current state according to $A^{(\ell)}(t)$, \emph{then} each agent averages the resulting values across all layers:
\begin{equation*}
x_i^{(\ell)}(t+1) = \frac{1}{m} \sum_{\ell=1}^m \sum_{j=1}^n A_{ij}^{(\ell)}(t) \, x_j^{(\ell)}(t), \quad i\in [n],\, \ell\in [m],\, t\ge 0.
\end{equation*}
Since $x_i(t)=x_i^{(1)}(t)=\dots=x_i^{(m)}(t)$ for all $i\in [n]$ and $t\ge 1$, it suffices to consider
\begin{equation}\label{eq:distribute_then_average}
x_i(t+1)= \frac{1}{m} \sum_{\ell=1}^m \sum_{j=1}^n A_{ij}^{(\ell)}(t) \, x_j(t), \quad i\in [n],\, t\ge 1.
\end{equation}

\paragraph{Mechanism II: Average-then-Distribute.}  
Each agent first computes the average of their neighbors’ states across all layers, and then redistributes within each layer according to $A^{(\ell)}(t)$:
\begin{equation}\label{eq:average_then_distribute}
x_i^{(\ell)}(t+1) = \sum_{j=1}^n A_{ij}^{(\ell)}(t) \left( \frac{1}{m} \sum_{k=1}^m x_j^{(k)}(t) \right), \quad i\in [n], \, \ell\in [m],\, t\ge 0.
\end{equation}

Observe that both mechanisms are \emph{conservative}, i.e., the total state across layers is preserved:
\[
\sum_{i\in [n],\ell\in [m]} x_i^{(\ell)}(t+1) = \sum_{i\in [n],\ell\in [m]} x_i^{(\ell)}(t).
\]

To study stability and convergence, we introduce quadratic functionals:
\begin{align*}
Z_t &= \sum_{i\in [n]} x_i(t)^2, \quad \text{for Mechanism I},  \\
W_t &= \sum_{\ell\in [m],i\in [n]} x_i^{(\ell)}(t)^2, \quad \text{for Mechanism II}.
\end{align*}

We now turn to an application of the proposed framework, namely the multilayer garbage disposal game.

\section{The multilayer garbage disposal game}

The multilayer garbage disposal game is an extension of the single-layer garbage disposal game. The term \emph{garbage} refers to undesirable resources, and each such resource is quantified by a nonnegative real number representing its magnitude. In the single-layer garbage disposal game, each agent offloads its garbage onto other agents at each time step. 

In the multilayer setting with $m$ layers and $n$ agents, $x_i^{(\ell)}(t)\ge 0$ denotes the amount of garbage held by agent $i$ in layer $\ell$ at time $t$. Each layer represents a type of social relationship, such as friendship or counterparty interaction. We represent the social structure in each layer using a graph. We now introduce the following mechanisms.

\subsection{Model Description}

Let $G^{(\ell)}(t)=([n],E^{(\ell)}(t))$ be a simple undirected graph representing layer $\ell$ at time $t$, with vertex set $[n]$ and edge set $E^{(\ell)}(t)$. Define
\[
\bar{x}_i(t)=\frac{1}{m}\sum_{k\in[m]}x_i^{(k)}(t)
\]
as the average amount of garbage held by agent $i$ across all layers at time $t$. For each layer $\ell$, let
\[
N_i^{(\ell)}(t)=\{k:(k,i)\in E^{(\ell)}(t)\}
\]
denote the neighborhood of agent $i$ in layer $\ell$ at time $t$.

We consider two redistribution mechanisms on the multilayer network: a \emph{distribute-then-average} mechanism and an \emph{average-then-distribute} mechanism.

In the distribute-then-average mechanism, each layer first performs a local redistribution of garbage, after which the resulting states are averaged across all layers. In this mechanism, each agent transfers a fraction $1/|E^{(\ell)}(t)|$ of its garbage to each of its neighbors in layer $\ell$. Consequently, agent $i$ transfers a total fraction
\[
\frac{|N_i^{(\ell)}(t)|}{|E^{(\ell)}(t)|}
\]
of its state and retains the remaining fraction
\[
1-\frac{|N_i^{(\ell)}(t)|}{|E^{(\ell)}(t)|}.
\]

Since $x_i^{(\ell)}(t)=x_i(t)$ for all $t\ge 1$, $i\in[n]$, and $\ell\in[m]$, this yields the update rule
\begin{equation}\label{Model:multilayerGDG DistributeThenAverage}
x_i(t+1)
=
\frac{1}{m}\sum_{\ell\in[m]}
\left[
x_i(t)\left(1-\frac{|N_i^{(\ell)}(t)|}{|E^{(\ell)}(t)|}\right)
+
\frac{1}{|E^{(\ell)}(t)|}\sum_{j\in N_i^{(\ell)}(t)} x_j(t)
\right],
\end{equation}
for all $t \ge 1$ and $i\in[n]$.

In the average-then-distribute mechanism, agents first compute the cross-layer average state $\bar{x}_i(t)$ and then apply the same redistribution rule within each layer using these averaged values. This leads to
\begin{equation}\label{Model:multilayerGDG AverageThenDistribute}
x_i^{(\ell)}(t+1)
=
\bar{x}_i(t)
\left(
1-\frac{|N_i^{(\ell)}(t)|}{|E^{(\ell)}(t)|}
\right)
+
\frac{1}{|E^{(\ell)}(t)|}
\sum_{j\in N_i^{(\ell)}(t)} \bar{x}_j(t),
\end{equation}
for all $t \ge 0$, $i\in[n]$, and $\ell\in[m]$.

Both~\eqref{Model:multilayerGDG DistributeThenAverage} and~\eqref{Model:multilayerGDG AverageThenDistribute} give rise to the same interaction matrix, namely
\[
A_{ii}^{(\ell)}(t)=1-\frac{|N_i^{(\ell)}(t)|}{|E^{(\ell)}(t)|}, 
\qquad
A_{ij}^{(\ell)}(t)=\frac{\mathbbm{1}\{j\in N_i^{(\ell)}(t)\}}{|E^{(\ell)}(t)|}, \quad i\neq j.
\]
In particular, \(A^{(\ell)}(t)\) is symmetric and doubly stochastic. Observe that \(A_{ii}^{(\ell)}(t)=0\) whenever \(G^{(\ell)}(t)\) is a star graph with node \(i\) as the center. In the special case of a single-layer two-agent system with connected \(G^{(\ell)}(t)\), the dynamics reduce to an exchange process. Moreover, for all \(n \neq 2\) and connected \(G^{(\ell)}(t)\), at most one vertex in \(G(A^{(\ell)}(t))\) does not have a loop.

\section{Main Theoretical Results}

We establish consensus results for Mechanisms~\eqref{eq:distribute_then_average} and~\eqref{eq:average_then_distribute}. The analysis relies on connectivity properties of the induced interaction graphs and conservation of total mass. The results are formulated in terms of recurrent connectivity along subsequences of time at which the induced interaction graphs admit a common subgraph with uniform positivity. In each case, convergence is guaranteed along such subsequences under which the interaction structure is sufficiently connected. The two mechanisms differ only in the structure of their associated time-varying stochastic matrices. Under these general conditions, both mechanisms converge to the global average of the initial states. We then specialize these results to the multilayer garbage disposal dynamics, where the assumptions reduce to more transparent connectivity requirements on the underlying layer graphs corresponding to Mechanisms~\eqref{Model:multilayerGDG DistributeThenAverage} and~\eqref{Model:multilayerGDG AverageThenDistribute}.

\begin{theorem}\label{Thm:ConsensusUnderM1}
In Mechanism~\eqref{eq:distribute_then_average}, assume that $\mathbb{E}[Z_0] < \infty$ and that there exists an increasing sequence $(t_k)_{k \ge 0} \subset \mathbb{Z}^+$ and a subgraph $H$ of $G(\bar{A}(t_k))$ for all $k \ge 0$ such that
\begin{itemize}
    \item $M(H)$ is connected,
    \item $\displaystyle \inf_{k \ge 0} \min_{(i,j) \in E(H)} \bar{A}_{ij}(t_k) > 0$,
    \item $\displaystyle \bigcup_{i \in V(H)} N_{H,i} = [n]$.
\end{itemize}
Then $x_i(t)$ converges to
\[
\frac{1}{n}\sum_{j \in [n]} x_j(0)
\]
as $t \to \infty$ for all $i \in [n]$.
\end{theorem}

Note that the conclusion of Theorem~\ref{Thm:ConsensusUnderM1} is equivalent to the statement that $x_i^{(\ell)}(t)$ converges to
\[
\frac{1}{mn}\sum_{p \in [m],\, j \in [n]} x_j^{(p)}(0)
\]
as $t \to \infty$ for all $i \in [n]$ and $\ell \in [m]$.

\begin{theorem}\label{Thm:ConsensusUnderM2}
In Mechanism~\eqref{eq:average_then_distribute}, assume that $\mathbb{E}[W_0] < \infty$ and that there exist $(\ell_k)_{k\ge0}\subset [m]$, an increasing sequence $(t_k)_{k \ge 0} \subset \mathbb{Z}^+$, and a subgraph $H$ of $G(A^{(\ell_k)}(t_k))$ for all $k \ge 0$ such that
\begin{itemize}
    \item $M(H)$ is connected,
    \item $\displaystyle \inf_{k \ge 0} \min_{(i,j) \in E(H)} A_{ij}^{(\ell_k)}(t_k) > 0$,
    \item $\displaystyle \bigcup_{i \in V(H)} N_{H,i} = [n]$.
\end{itemize}
Then $x_i^{(\ell)}(t)$ converges to
\[
\frac{1}{mn} \sum_{p \in [m],\, j \in [n]} x_j^{(p)}(0)
\]
as $t \to \infty$ for all $i \in [n]$ and $\ell \in [m]$.
\end{theorem}

The following corollaries specialize the preceding results to the multilayer garbage disposal game, where the interaction matrices $A^{(\ell)}(t)$ are induced by simple undirected graphs $G^{(\ell)}(t)$.

\begin{corollary}\label{Cor:ConsensusUnderDistributeThenAverage}
In Mechanism~\eqref{Model:multilayerGDG DistributeThenAverage}, assume that $\mathbb{E}[Z_0] < \infty$ and that $\bigcup_{\ell\in [m]}G^{(\ell)}(t)$ is connected infinitely often. Then, for all $n \neq 2$, $x_i(t)$ converges to
\[
\frac{1}{n}\sum_{j \in [n]} x_j(0)
\]
as $t \to \infty$ for all $i \in [n]$.
\end{corollary}

\begin{corollary}\label{Cor:ConsensusUnderAverageThenDistribute}
In Mechanism~\eqref{Model:multilayerGDG AverageThenDistribute}, assume that $\mathbb{E}[W_0] < \infty$ and that there exist $(\ell_k)_{k\ge 0}\subset [m]$ and an increasing sequence $(t_k)_{k\ge 0}\subset \mathbb{Z}^+$ such that $G^{(\ell_k)}(t_k)$ is connected for all $k\ge 0$. Then, for all $n \neq 2$, $x_i^{(\ell)}(t)$ converges to
\[
\frac{1}{mn} \sum_{p \in [m],\, j \in [n]} x_j^{(p)}(0)
\]
as $t \to \infty$ for all $i \in [n]$ and $\ell \in [m]$.
\end{corollary}

\section{Theoretical Analysis}

In this section, we develop the technical tools required to establish convergence of the multilayer dynamics. We first derive key dissipation identities based on quadratic Lyapunov functionals, which quantify how stochastic averaging reduces dispersion over time. These identities yield supermartingale structures that guarantee almost sure convergence under minimal integrability assumptions. We then analyze each mechanism separately, showing how connectivity of induced interaction graphs drives asymptotic consensus. Finally, we introduce a set of structural graph lemmas that will be used to translate the abstract connectivity conditions in the main theorems into simpler conditions for the multilayer garbage disposal model.

\subsection{Common Preliminary Lemmas}

We begin by establishing a fundamental quadratic decomposition identity that underlies all subsequent Lyapunov arguments. This identity characterizes the variance reduction induced by stochastic averaging and will be repeatedly used to express the evolution of quadratic functionals in terms of pairwise disagreement terms. It serves as the basic analytic tool for deriving dissipation formulas in both multilayer mechanisms.

\begin{lemma}[Quadratic Dissipation Identity]\label{Lemma:QuadraticDissipationIdentity}
For any stochastic vector $(a_1,\ldots,a_n)$ and any $y_1,\ldots,y_n \in \mathbb{R}$,
\[
\Bigl(\sum_{i\in [n]} a_i y_i\Bigr)^2
=
\sum_{i\in [n]} a_i y_i^2
-
\frac12 \sum_{i,j\in [n]} a_i a_j (y_i-y_j)^2.
\]
\end{lemma}

\begin{proof}
We compute
\begin{align*}
\Bigl(\sum_{i\in [n]} a_i y_i\Bigr)^2
&= \sum_{i,j\in [n]} a_i a_j y_i y_j \\
&= \sum_{i,j\in [n]} \frac12 a_i a_j \bigl[y_i^2 + y_j^2 - (y_i-y_j)^2\bigr] \\
&= \frac12 \sum_{i,j\in [n]} a_i a_j (y_i^2+y_j^2)
   - \frac12 \sum_{i,j\in [n]} a_i a_j (y_i-y_j)^2.
\end{align*}

Observe that
\begin{align*}
\frac12 \sum_{i,j\in [n]} a_i a_j (y_i^2+y_j^2)
&= \frac12 \sum_{i,j\in [n]} a_i a_j y_i^2
 + \frac12 \sum_{i,j\in [n]} a_i a_j y_j^2 \\
&= \frac12 \sum_{i,j\in [n]} a_i a_j y_i^2
 + \frac12 \sum_{i,j\in [n]} a_j a_i y_i^2 \\
&= \sum_{i,j\in [n]} a_i a_j y_i^2 \\
&= \Bigl(\sum_{j\in [n]} a_j\Bigr)\Bigl(\sum_{i\in [n]} a_i y_i^2\Bigr) \\
&= \sum_{i\in [n]} a_i y_i^2.
\end{align*}

Therefore,
\[
\Bigl(\sum_{i\in [n]} a_i y_i\Bigr)^2
=
\sum_{i\in [n]} a_i y_i^2
-
\frac12 \sum_{i,j\in [n]} a_i a_j (y_i-y_j)^2.
\]
\end{proof}

\subsection{Mechanism I: Distribute then Average}

We now analyze the distribute-then-average mechanism. The key step is to express the evolution of the quadratic functional in terms of the averaged interaction matrix and to identify a dissipation term that quantifies the decay of disagreement across agents. This leads to a nonnegative supermartingale structure, which is then used to establish convergence under recurrent connectivity of the induced interaction graphs.

\begin{lemma}\label{Lemma:SupermartingaleZ_t}
The process $(Z_t)_{t\ge 1}$ is nonincreasing and satisfies
\[
Z_t-Z_{t+1}
=
\frac12 \sum_{i,j,k\in [n]} \bar{A}_{ij}(t)\bar{A}_{ik}(t)(x_j(t)-x_k(t))^2,
\]
where
\[
\bar{A}_{ij}(t) = \frac1m \sum_{\ell \in [m]} A_{ij}^{(\ell)}(t), \quad i,j \in [n].
\]
\end{lemma}

\begin{proof}
Let
\[
x_i^\star = x_i(t+1), \qquad x_i = x_i(t), \qquad \bar{A}_{ij} = \bar{A}_{ij}(t).
\]
Then
\[
x_i^{\star 2}
=
\left(
\frac{1}{m} \sum_{\ell\in [m]} \sum_{j\in [n]} A_{ij}^{(\ell)} x_j
\right)^2
=
\left(
\sum_{j\in [n]} \bar{A}_{ij} x_j
\right)^2,
\quad \text{for all } i \in [n].
\]

Since each matrix $A^{(k)}$ is doubly stochastic, their average $\bar{A}_{ij} = \frac{1}{m} \sum_{\ell\in [m]} A_{ij}^{(\ell)}$ is also doubly stochastic.  
Therefore,
\[
\sum_{i\in [n]} \bar{A}_{ij} = \sum_{j\in [n]} \bar{A}_{ij} = 1.
\]

By Lemma~\ref{Lemma:QuadraticDissipationIdentity}, we have
\begin{align*}
Z_{t+1} &= \sum_{i\in [n]} x_i^{\star 2} \\
&= \sum_{i\in [n]} \left(\sum_{j\in [n]} \bar{A}_{ij} x_j\right)^2 \\
&= \sum_{i\in [n]} \Biggl[ \sum_{j\in [n]} \bar{A}_{ij} x_j^2
- \frac12 \sum_{j,k\in [n]} \bar{A}_{ij} \bar{A}_{ik} (x_j - x_k)^2 \Biggr] \\
&= \sum_{i,j\in [n]} \bar{A}_{ij} x_j^2
- \frac12 \sum_{i,j,k\in [n]} \bar{A}_{ij} \bar{A}_{ik} (x_j - x_k)^2 \\
&= \sum_{j\in [n]} \left( \sum_{i\in [n]} \bar{A}_{ij} \right) x_j^2
- \frac12 \sum_{i,j,k\in [n]} \bar{A}_{ij} \bar{A}_{ik} (x_j - x_k)^2 \\
&= \sum_{j\in [n]} x_j^2
- \frac12 \sum_{i,j,k\in [n]} \bar{A}_{ij} \bar{A}_{ik} (x_j - x_k)^2 \\
&= Z_t - \frac12 \sum_{i,j,k\in [n]} \bar{A}_{ij} \bar{A}_{ik} (x_j - x_k)^2.
\end{align*}

Hence,
\[
Z_t - Z_{t+1} = \frac12 \sum_{i,j,k\in [n]} \bar{A}_{ij} \bar{A}_{ik} (x_j - x_k)^2.
\]
\end{proof}

\begin{lemma}\label{Lemma:LimitOfZ_t}
Let $\mathbb{E}[Z_0] < \infty$. Then the process $(Z_t)_{t \ge 1}$ converges almost surely to a random variable $Z$ with finite expectation as $t \to \infty$.
\end{lemma}

\begin{proof}
It follows from the assumption and Lemma~\ref{Lemma:SupermartingaleZ_t} that $(Z_t)_{t \ge 1}$ is a nonnegative supermartingale. Hence, by the martingale convergence theorem, $Z_t$ converges almost surely to a random variable $Z$ with finite expectation as $t \to \infty$.
\end{proof}

\begin{proof}[\bf Proof of Theorem~\ref{Thm:ConsensusUnderM1}]
    Let $\delta = \inf_{k \ge 0} \min_{(i,j) \in E(H)} \bar{A}_{ij}(t_k).$ Then $\delta > 0$ by assumption. Since $x_i^{(\ell)}(t)$ is a convex combination of $(x_j^{(s)}(t))_{s \in [m],\, j \in [n]}$, it suffices to show that
\[
x_i(t_s) \to \frac{1}{mn} \sum_{\ell \in [m],\, j \in [n]} x_j^{(\ell)}(0)
\]
as $s \to \infty$ for all $i \in [n]$.

For Mechanism~\eqref{eq:distribute_then_average}, it follows from Lemma~\ref{Lemma:SupermartingaleZ_t} that
\[
Z_{t_s} - Z_{t_s+1} \ge \frac{1}{2} \sum_{i \in [n],\, j,k \in N_{H,i}} \delta^2 \big(x_j(t_s) - x_k(t_s)\big)^2.
\]
By Lemma~\ref{Lemma:LimitOfZ_t}, we have $Z_{t_s} - Z_{t_s+1} \to 0$ as $s \to \infty$. Hence,
\[
x_j(t_s) - x_k(t_s) \to 0
\]
as $s \to \infty$ for all $i \in [n]$ and $j,k \in N_{H,i}$.

Since $\bigcup_{i \in V(H)} N_{H,i} = [n]$, for any $i_1, i_2 \in [n]$, there exist $\hat{i}_1, \hat{i}_2 \in V(H)$ such that $i_1 \in N_{H,\hat{i}_1}$ and $i_2 \in N_{H,\hat{i}_2}$. Since $M(H)$ is connected, there exists a path from $\hat{i}_1$ to $\hat{i}_2$, say
\[
\hat{i}_1 = c_0 - c_1 - \cdots - c_r = \hat{i}_2.
\]
For each $q \in [r]$, pick $j_q \in N_{H,c_{q-1}} \cap N_{H,c_q}$. Then
\[
x_{i_1}(t_s) - x_{i_2}(t_s)
= \big(x_{i_1}(t_s) - x_{j_1}(t_s)\big)
+ \sum_{q \in [r-1]} \big(x_{j_q}(t_s) - x_{j_{q+1}}(t_s)\big)
+ \big(x_{j_r}(t_s) - x_{i_2}(t_s)\big),
\]
where $i_1, j_1 \in N_{H,\hat{i}_1}$, $j_q, j_{q+1} \in N_{H,c_q}$ for all $q \in [r-1]$, and $j_r, i_2 \in N_{H,\hat{i}_2}$. Therefore,
\[
x_{i_1}(t_s) - x_{i_2}(t_s) \to 0
\]
as $s \to \infty$.

Since $(x_i(t))_{i \in [n]}$ is conserved,
\[
\frac{1}{mn} \sum_{\ell \in [m],\, i \in [n]} x_i^{(\ell)}(0) - x_{i_2}(t_s)
= \frac{1}{n} \sum_{i_1 \in [n]} x_{i_1}(t_s) - x_{i_2}(t_s) \to 0
\]
as $s \to \infty$ for all $i_2 \in [n]$.

\end{proof}

\subsection{Mechanism II: Average then Distribute}

We next study the average-then-distribute mechanism, where coupling across layers occurs prior to local redistribution. This additional averaging step introduces a two-level dissipation structure: one term capturing inter-layer synchronization and another capturing intra-layer mixing. We show that both effects jointly yield a supermartingale evolution of the corresponding Lyapunov functional, leading again to almost sure convergence under appropriate connectivity conditions.

\begin{lemma}\label{Lemma:SupermartingaleW_t}
The process $(W_t)_{t \ge 0}$ is nonincreasing and satisfies
\begin{align*}
W_t-W_{t+1}
&=
\sum_{\ell\in[m],\,j\in[n]}
\left(x_j^{(\ell)}(t)-\bar{x}_j(t)\right)^2 \\
&\quad+
\frac12
\sum_{\ell\in[m],\,i,j,k\in[n]}
A_{ij}^{(\ell)}(t)A_{ik}^{(\ell)}(t)
\left(\bar{x}_j(t)-\bar{x}_k(t)\right)^2.
\end{align*}
where
\[
\bar{x}_j(t)=\frac{1}{m}\sum_{\ell \in [m]} x_j^{(\ell)}(t), \qquad j \in [n].
\]
\end{lemma}

\begin{proof}
Let
\[
x_i^{(\ell)} = x_i^{(\ell)}(t),
\qquad
A_{ij}^{(\ell)} = A_{ij}^{(\ell)}(t).
\]
Observe that
\begin{align*}
W_{t+1}
&=
\sum_{\ell\in [m],\, i\in [n]} x_i^{(\ell)}(t+1)^2 \\
&=
\sum_{\ell\in [m],\, i\in [n]}
\left(\sum_{j\in [n]} A_{ij}^{(\ell)} \bar{x}_j\right)^2 \\
&=
\sum_{\ell\in [m],\, i\in [n]}
\left[
\sum_{j\in [n]} A_{ij}^{(\ell)} \bar{x}_j^2
-
\frac12
\sum_{j,k\in [n]}
A_{ij}^{(\ell)} A_{ik}^{(\ell)}
(\bar{x}_j-\bar{x}_k)^2
\right] \\
&=
\sum_{\ell\in [m],\, j\in [n]}
\left(\sum_{i\in [n]} A_{ij}^{(\ell)}\right)\bar{x}_j^2
-
\frac12
\sum_{\ell\in [m],\, i,j,k\in [n]}
A_{ij}^{(\ell)} A_{ik}^{(\ell)}
(\bar{x}_j-\bar{x}_k)^2 \\
&=
m \sum_{j\in [n]} \bar{x}_j^2
-
\frac12
\sum_{\ell\in [m],\, i,j,k\in [n]}
A_{ij}^{(\ell)} A_{ik}^{(\ell)}
(\bar{x}_j-\bar{x}_k)^2.
\end{align*}

On the other hand,
\begin{align*}
W_t
&=
\sum_{\ell\in [m],\, j\in [n]} \left(x_j^{(\ell)}\right)^2 \\
&=
\sum_{\ell\in [m],\, j\in [n]}
\left[
\bar{x}_j^2
+
\left(x_j^{(\ell)}-\bar{x}_j\right)^2
+
2\bar{x}_j\left(x_j^{(\ell)}-\bar{x}_j\right)
\right] \\
&=
m\sum_{j\in [n]} \bar{x}_j^2
+
\sum_{\ell\in [m],\, j\in [n]}
\left(x_j^{(\ell)}-\bar{x}_j\right)^2
+
2\sum_{j\in [n]}
\bar{x}_j
\sum_{\ell\in [m]}
\left(x_j^{(\ell)}-\bar{x}_j\right) \\
&=
m\sum_{j\in [n]} \bar{x}_j^2
+
\sum_{\ell\in [m],\, j\in [n]}
\left(x_j^{(\ell)}-\bar{x}_j\right)^2,
\end{align*}
since
\[
\sum_{\ell\in [m]} \left(x_j^{(\ell)}-\bar{x}_j\right)=0.
\]

Thus,
\[
W_t-W_{t+1}
=
\sum_{\ell\in [m],\, j\in [n]}
\left(x_j^{(\ell)}-\bar{x}_j\right)^2
+
\frac12
\sum_{\ell\in [m],\, i,j,k\in [n]}
A_{ij}^{(\ell)} A_{ik}^{(\ell)}
(\bar{x}_j-\bar{x}_k)^2.
\]
\end{proof}

\begin{lemma}\label{Lemma:LimitOfW_t}
Let $\mathbb{E}[W_0] < \infty$. Then the process $(W_t)_{t \ge 0}$ converges almost surely to a random variable $W$ with finite expectation as $t \to \infty$.
\end{lemma}

\begin{proof}
It follows from the assumption and Lemma~\ref{Lemma:SupermartingaleW_t} that $(W_t)_{t \ge 0}$ is a nonnegative supermartingale. Hence, by the martingale convergence theorem, $W_t$ converges almost surely to a random variable $W$ with finite expectation as $t \to \infty$.
\end{proof}

\begin{proof}[\bf Proof of Theorem~\ref{Thm:ConsensusUnderM2}]
Let
\[
\delta = \inf_{k \ge 0} \min_{(i,j) \in E(H)} A_{ij}^{(\ell_k)}(t_k).
\]
Then $\delta > 0$ by assumption. Since $x_i^{(\ell)}(t)$ is a convex combination of $(x_j^{(s)}(t))_{s \in [m],\, j \in [n]}$, it suffices to show that
\[
x_i^{(k)}(t_s) \to \frac{1}{mn} \sum_{\ell \in [m],\, j \in [n]} x_j^{(\ell)}(0)
\]
as $s \to \infty$ for all $i \in [n]$ and $k \in [m]$.

For Mechanism~\eqref{eq:average_then_distribute}, it follows from Lemma~\ref{Lemma:SupermartingaleW_t} that
\begin{align*}
W_{t_s}-W_{t_s+1}
&\ge
\sum_{\ell\in[m],\,j\in[n]}
\left(x_j^{(\ell)}(t_s)-\bar{x}_j(t_s)\right)^2 \\
&\quad+
\frac12
\sum_{i \in [n],\, j,k \in N_{H,i}}
\delta^2
\left(\bar{x}_j(t_s)-\bar{x}_k(t_s)\right)^2.
\end{align*}

By Lemma~\ref{Lemma:LimitOfW_t}, we have $W_{t_s} - W_{t_s+1} \to 0$ as $s \to \infty$. Hence,
\[
x_j^{(\ell)}(t_s)-\bar{x}_j(t_s) \to 0
\]
for all $\ell\in [m]$ and $j\in [n]$, and
\[
\bar{x}_j(t_s)-\bar{x}_k(t_s)\to 0
\]
for all $i \in [n]$ and $j,k \in N_{H,i}$ as $s \to \infty$.

Since $\bigcup_{i \in V(H)} N_{H,i} = [n]$, for any $i_1, i_2 \in [n]$, there exist $\hat{i}_1, \hat{i}_2 \in V(H)$ such that $i_1 \in N_{H,\hat{i}_1}$ and $i_2 \in N_{H,\hat{i}_2}$. Since $M(H)$ is connected, there exists a path from $\hat{i}_1$ to $\hat{i}_2$, say
\[
\hat{i}_1 = c_0 - c_1 - \cdots - c_r = \hat{i}_2.
\]
For each $q \in [r]$, pick $j_q \in N_{H,c_{q-1}} \cap N_{H,c_q}$. Then
\begin{align*}
x_{i_1}^{(\ell_1)}(t_s) - x_{i_2}^{(\ell_2)}(t_s) &=
\big(x_{i_1}^{(\ell_1)}(t_s)-\bar{x}_{i_1}(t_s)\big)
+\big(\bar{x}_{i_1}(t_s) - \bar{x}_{j_1}(t_s)\big) \\
&\quad+ \sum_{q \in [r-1]}
\big(\bar{x}_{j_q}(t_s) - \bar{x}_{j_{q+1}}(t_s)\big) \\
&\quad+
\big(\bar{x}_{j_r}(t_s)-\bar{x}_{i_2}(t_s)\big)
+\big(\bar{x}_{i_2}(t_s) - x_{i_2}^{(\ell_2)}(t_s)\big),
\end{align*}
where $i_1, j_1 \in N_{H,\hat{i}_1}$, $j_q, j_{q+1} \in N_{H,c_q}$ for all $q \in [r-1]$, and $j_r, i_2 \in N_{H,\hat{i}_2}$. Therefore,
\[
x_{i_1}^{(\ell_1)}(t_s) - x_{i_2}^{(\ell_2)}(t_s) \to 0
\]
as $s \to \infty$.

Since $\big(x_i^{(\ell)}(t)\big)_{\ell \in [m],\, i \in [n]}$ preserves the average,
\[
\frac{1}{mn} \sum_{\ell \in [m],\, i \in [n]} x_i^{(\ell)}(0)
- x_{i_2}^{(\ell_2)}(t_s)
=
\frac{1}{mn}
\sum_{i_1 \in [n],\, \ell_1\in [m]}
\left(
x_{i_1}^{(\ell_1)}(t_s) - x_{i_2}^{(\ell_2)}(t_s)
\right)
\to 0
\]
as $s \to \infty$ for all $i_2 \in [n]$ and $\ell_2\in [m]$.
\end{proof}

\subsection{Connectivity Lemmas and Technical Preliminaries}

We collect several graph-theoretic lemmas that connect the structural properties of the auxiliary graph $M(G)$ with standard connectivity of $G$. These results are used to verify the connectivity assumptions in the main theorems under the multilayer garbage disposal model. In particular, they allow us to translate conditions on induced interaction graphs into simpler conditions on the underlying layer graphs, thereby enabling the derivation of the corollaries.

\begin{lemma}
If $G$ is an undirected graph such that $M(G)$ is connected, then $G$ is connected.
\end{lemma}

\begin{proof}
Since $M(G)$ is connected, for any $i,j\in [n]$, there exists an $i,j$-path in $M(G)$, say
\[
i=k_0-k_1-\cdots-k_s=j.
\]
Choose $\hat{k}_r\in N_{G,k_{r-1}}\cap N_{G,k_r}$ for each $r\in [s]$. Then
\[
i=k_0-\hat{k}_1-k_1-\hat{k}_2-k_2-\cdots-\hat{k}_s-k_s=j
\]
is an $i,j$-path in $G$. Hence, $G$ is connected.
\end{proof}

\begin{lemma}\label{Lemma:UnionOfIndividualNeighborhoodEqualsPopulation}
If $G$ is an undirected graph in which every vertex has degree at least 1, then
\[
\bigcup_{i\in [n]} N_{G,i} = [n].
\]
\end{lemma}

\begin{proof}
It follows from the assumption that for each $j \in [n]$, there exists $i \in [n]$ such that $(j,i) \in E(G)$, which implies $j \in N_{G,i}$. Therefore,
\[
[n] \subseteq \bigcup_{i\in [n]} N_{G,i}.
\]
The reverse inclusion follows immediately from the definition of the neighborhoods $N_{G,i}$, $i \in [n]$.
\end{proof}

\begin{lemma}\label{Lemma:ConditionsForM(G)Connected}
If $G$ is an undirected connected graph such that, for every edge $(i,j)\in E(G)$, at least one of the vertices $i$ and $j$ contains a loop, then $M(G)$ is connected.
\end{lemma}

\begin{proof}
It follows from the assumption that, for every $(i,j)\in E(G)$,
\[
N_{G,i}\cap N_{G,j}\neq\emptyset.
\]
Hence, $(i,j)\in E(M(G))$ for every $(i,j)\in E(G)$. Therefore,
\[
E(G)\subseteq E(M(G)).
\]
Since $G$ is connected, it follows that $M(G)$ is connected.
\end{proof}

\begin{proof}[\bf Proof of Corollary~\ref{Cor:ConsensusUnderDistributeThenAverage}]
It is clear that the result holds for $n=1$. We claim that the result also holds for $n>2$. By the assumption and the finiteness of the social graphs, there exist $(t_k)_{k\ge 0} \subset \mathbb{Z}^+$ and a connected graph $H$ such that
\[
G(\bar{A}(t_k)) = H \quad \text{for all } k \ge 0.
\]
Under Mechanism~\eqref{Model:multilayerGDG DistributeThenAverage}, $H$ has at most one vertex without a loop. Thus, by Lemma~\ref{Lemma:ConditionsForM(G)Connected}, $M(H)$ is connected. Together with Lemma~\ref{Lemma:UnionOfIndividualNeighborhoodEqualsPopulation}, it follows that
\[
\bigcup_{i\in [n]}N_{H,i}=[n].
\]
Moreover, we have
\[
\inf_{k \ge 0} \min_{(i,j) \in E(H)} \bar{A}_{ij}(t_k)
\ge \frac{1}{m\max_{\ell\in [m]}|E^{(\ell)}|}
\ge \frac{1}{m\binom{n}{2}}
>0.
\]
By Theorem~\ref{Thm:ConsensusUnderM1}, the result holds for $n>1$. This completes the proof.
\end{proof}

\begin{proof}[\bf Proof of Corollary~\ref{Cor:ConsensusUnderAverageThenDistribute}]
It is clear that the result holds for $n=1$. We claim that the result also holds for $n>2$. By the assumption and finiteness of the social graphs, there exist $(s_r)_{r\ge 0} \subset \mathbb{N}$ and a connected graph $H$ such that
\[
G(A^{(\ell_{s_r})}(t_{s_r})) = H \quad \text{for all } r \ge 0.
\]
Under Mechanism~\eqref{Model:multilayerGDG AverageThenDistribute}, $H$ has at most one vertex without a loop. Thus, by Lemma~\ref{Lemma:ConditionsForM(G)Connected}, $M(H)$ is connected. Together with Lemma~\ref{Lemma:UnionOfIndividualNeighborhoodEqualsPopulation}, it follows that
\[
\bigcup_{i\in [n]}N_{H,i}=[n].
\]
Moreover, we have
\[
\inf_{r \ge 0} \min_{(i,j) \in E(H)} A^{(\ell_{s_r})}_{ij}(t_{s_r})
\ge \frac{1}{\max_{\ell\in [m]}|E^{(\ell)}|}
\ge \frac{1}{\binom{n}{2}}
>0.
\]
By Theorem~\ref{Thm:ConsensusUnderM2}, the result holds for $n>1$. This completes the proof.
\end{proof}

\section{Numerical Illustrations}

We consider a network of $n$ agents, where $x_i^{(k)}(t) \in \mathbb{R}$ denotes the state of agent $i$ at discrete time $t$ in layer $k$. The state represents a generic scalar quantity, such as an opinion, belief, resource level, or load (e.g., garbage amount), depending on the application.

We illustrate the behavior of Mechanisms~\eqref{eq:distribute_then_average} and~\eqref{eq:average_then_distribute} through Monte Carlo simulations. Throughout all experiments, we set $n=100$ agents, $m=5$ layers (for the multilayer case), and horizon $T=10$. The initial states are independently sampled from a standard normal distribution.

\subsection{Mechanism I: Single-layer switching dynamics}

We first consider a single-layer opinion dynamics model under time-varying random interaction graphs. 
At each time step $t$, we construct an averaged stochastic matrix
\[
\bar{A}(t) = \frac{1}{m} \sum_{\ell=1}^{m} A^{(\ell)}(t),
\]
where each $A^{(\ell)}(t)$ is generated independently based on a \textbf{simple undirected graph} according to the following mechanism: with probability $0.5$, the underlying graph is a random spanning tree generated via a Prüfer sequence; otherwise, it is a sparse Erd\H{o}s--R\'enyi graph with edge probability $p=0.03$, which may be disconnected.

Each graph is transformed into a Metropolis weight matrix defined by
\[
A_{ij} = \frac{1}{1 + \max(d_i, d_j)}, \quad i \neq j,
\]
where $d_i$ denotes the degree of node $i$ in the corresponding graph, and the diagonal entries are chosen to ensure row-stochasticity. Since $A$ is symmetric, it follows that $A$ is doubly stochastic. Moreover,
\[
A_{ii}
= 1 - \sum_{j \neq i} A_{ij}
\ge 1 - \frac{d_i}{1+d_i}
= \frac{1}{1+d_i}
> 0,
\]
which implies that every node in $G(A)$ has a loop.

The opinion update rule is given by
\[
x(t+1) = \bar{A}(t) x(t).
\]

\begin{figure}[H]
\centering
\includegraphics[width=0.75\textwidth]{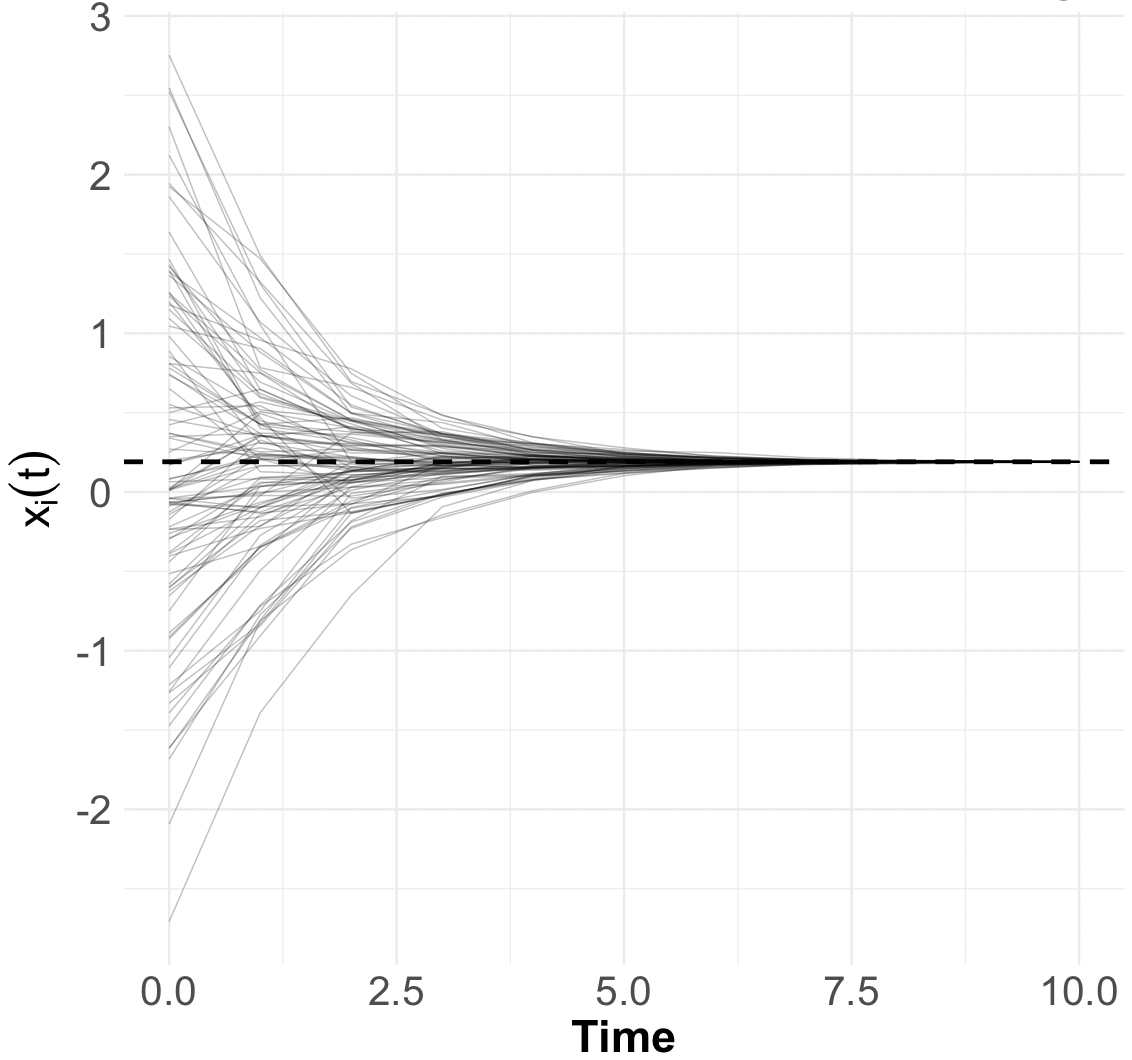}
\caption{Mechanism I: Evolution of opinions under random tree and sparse graph switching. Each curve represents an agent trajectory, and the dashed line indicates the initial empirical mean.}
\label{fig:M1}
\end{figure}

Figure~\ref{fig:M1} illustrates the evolution of agent opinions. Despite random switching between well-connected tree structures and sparse Erd\H{o}s--R\'enyi graphs (which may be disconnected), the system exhibits convergence toward consensus behavior.

\subsection{Mechanism II: Multilayer switching dynamics}

We next extend the model to a multilayer setting with $m=5$ interacting layers. Let $x_i^{(k)}(t)$ denote the opinion of agent $i$ in layer $k$. At each time step, we first compute the cross-layer average:
\[
\bar{x}_i(t) = \frac{1}{m} \sum_{k=1}^{m} x_i^{(k)}(t).
\]

For each layer $k$, an independent random graph is generated. With probability $0.5$, the graph is a random tree; otherwise, it is a sparse Erd\H{o}s--R\'enyi graph. Each graph is then mapped to a Metropolis matrix $A^{(k)}(t)$.

The layer-wise update is given by
\[
x^{(k)}(t+1) = A^{(k)}(t)\,\bar{x}(t),
\]
where \(x^{(k)}(t) = (x_1^{(k)}(t), \ldots, x_n^{(k)}(t))^T\) denotes the state vector of layer \(k\) at time \(t\), and \(\bar{x}(t) = (\bar{x}_1(t), \ldots, \bar{x}_n(t))^T\) is the vector of layer-averaged states.

\begin{figure}[H]
\centering
\includegraphics[width=0.75\textwidth]{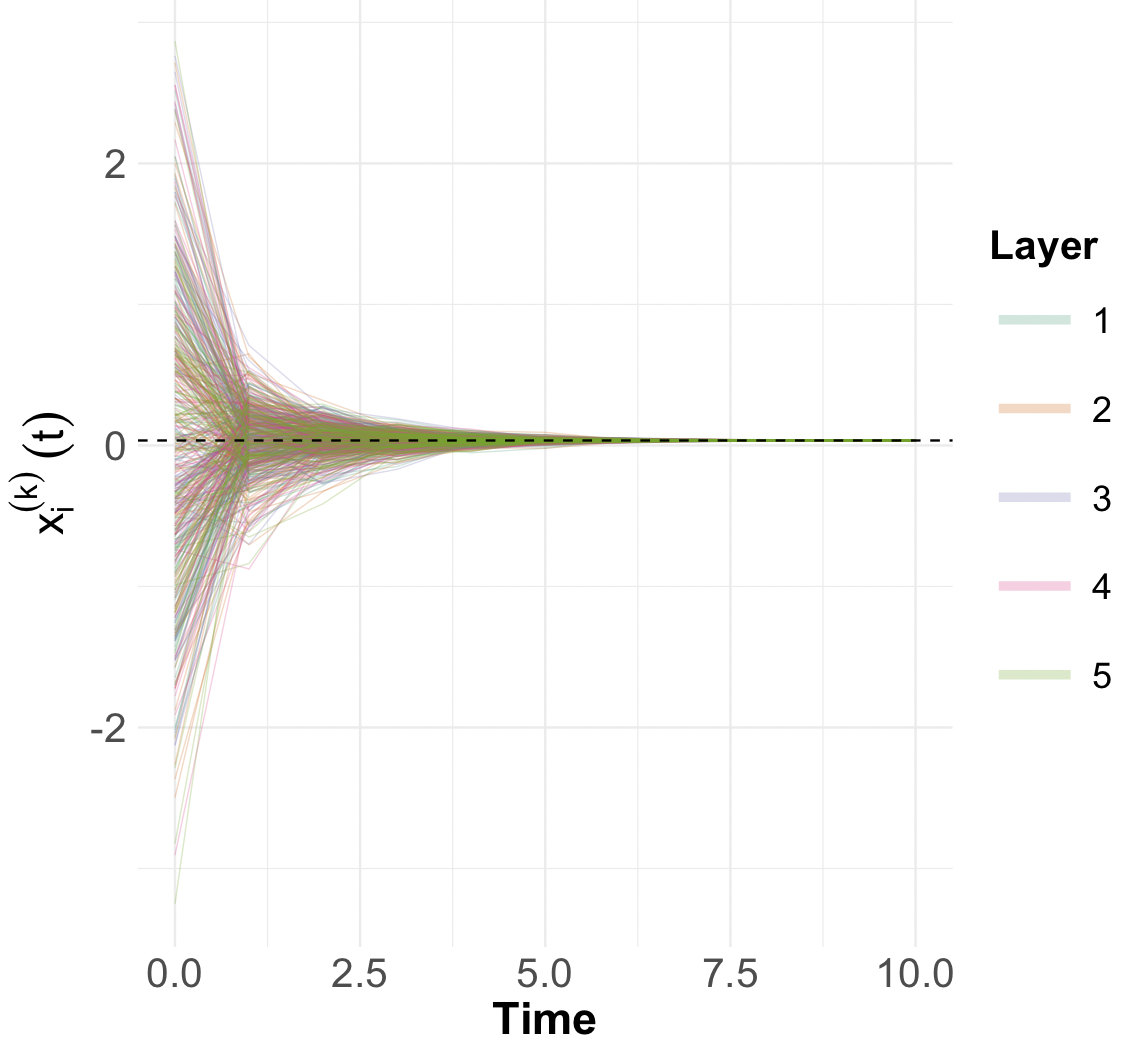}
\caption{Mechanism II: Multilayer opinion dynamics under switching random graphs. Different colors correspond to different layers. The dashed line indicates the initial global mean.}
\label{fig:M2}
\end{figure}

Figure~\ref{fig:M2} shows the multilayer evolution. Compared to the single-layer case, the additional layer coupling introduces richer transient heterogeneity while still preserving convergence behavior.

\subsection{Numerical illustrations for corollaries}

We provide Monte Carlo simulations to illustrate the convergence results established in Corollary~\ref{Cor:ConsensusUnderDistributeThenAverage} and Corollary~\ref{Cor:ConsensusUnderAverageThenDistribute}. The multilayer garbage disposal dynamics are implemented under Mechanisms~\eqref{Model:multilayerGDG DistributeThenAverage} and~\eqref{Model:multilayerGDG AverageThenDistribute}, respectively.

We consider a network of $n=100$ agents and $m=5$ layers. The simulation horizon is set to $T=400$ to clearly observe asymptotic behavior. The connectivity at each time step is generated according to a switching mechanism: with probability $0.15$, the underlying graph is a random spanning tree generated via a Prüfer sequence (ensuring connectivity), while with probability $0.85$, it is a sparse Erd\H{o}s--R\'enyi graph with edge probability $p=0.03$, which may be disconnected.

For each generated simple undirected graph $G^{(\ell)}(t)$, we construct the interaction matrix $A^{(\ell)}(t)$ according to the uniform redistribution rule:
\[
A_{ij}^{(\ell)}(t) =
\begin{cases}
\frac{1}{|E^{(\ell)}(t)|}, & i \neq j,\\[4pt]
1 - \frac{|N_i^{(\ell)}(t)|}{|E^{(\ell)}(t)|}, & i = j,\\[4pt]
0, & \text{otherwise},
\end{cases}
\]
where $E^{(\ell)}(t)$ denotes the edge set of $G^{(\ell)}(t)$ and
$N_i^{(\ell)}(t)=\{j : (j,i)\in E^{(\ell)}(t)\}$ denotes the neighborhood of node $i$ in $G^{(\ell)}(t)$.

Initial states are independently drawn from a standard normal distribution and projected onto $\mathbb{R}_+$ via truncation, ensuring non-negative initial conditions consistent with the interpretation of state variables as quantities such as resource or garbage levels. We report sample trajectories for both mechanisms.

\begin{figure}[H]
\centering
\includegraphics[width=0.7\textwidth]{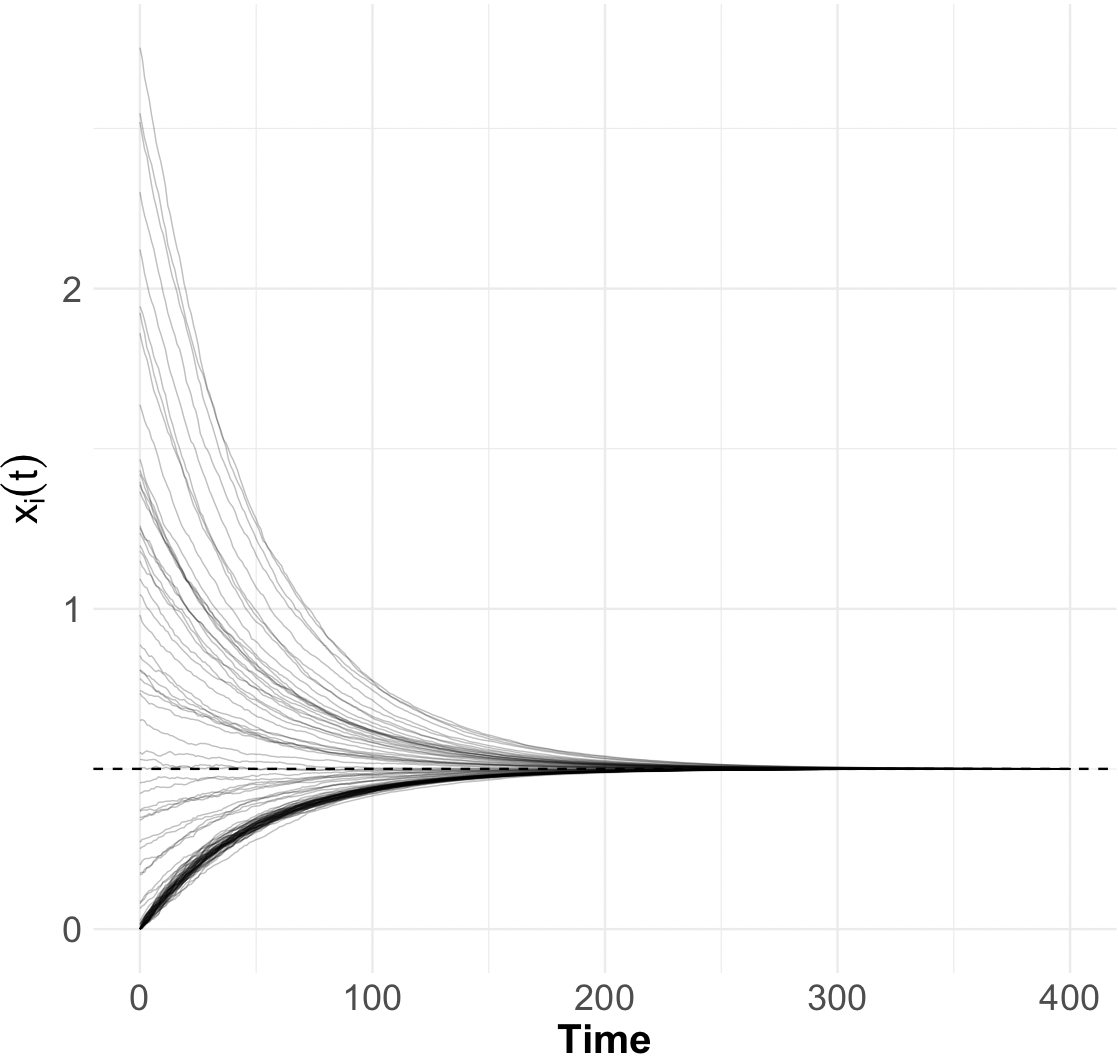}
\caption{Corollary simulation (Mechanism I): convergence of the single-layer garbage disposal game under intermittent connectivity. Each curve represents an agent trajectory.}
\label{fig:M1_cor}
\end{figure}

Figure~\ref{fig:M1_cor} illustrates the evolution under Mechanism I. Despite frequent sparse (possibly disconnected) graphs, the system converges to the initial empirical mean.

\begin{figure}[H]
\centering
\includegraphics[width=0.7\textwidth]{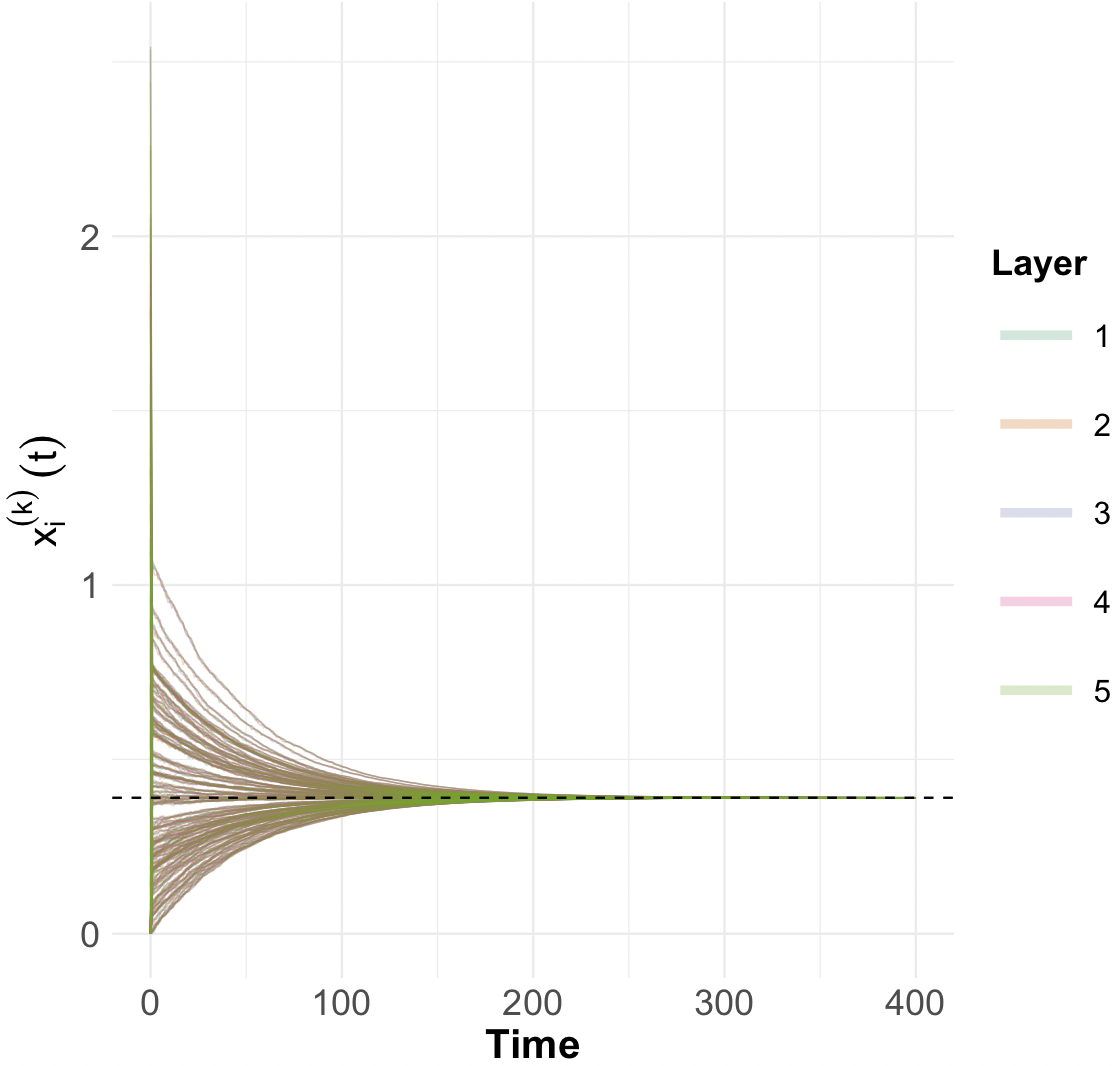}
\caption{Corollary simulation (Mechanism II): multilayer garbage disposal game dynamics under switching random graphs. Different colors correspond to different layers.}
\label{fig:M2_cor}
\end{figure}

Figure~\ref{fig:M2_cor} shows the multilayer evolution. The cross-layer coupling, combined with intermittent connectivity, leads to convergence toward the global average state while preserving transient heterogeneity across layers.

\subsection{Summary of numerical behavior}

Across the mechanisms, we observe that:

\begin{itemize}
\item The Metropolis construction and the uniform redistribution construction guarantee stochasticity of the update matrices.
\item Intermittent connectivity is sufficient to ensure global mixing, in the sense that information (or, in the multilayer garbage disposal game interpretation, the distributed garbage amount) propagates throughout the entire network over time, despite possible temporary disconnections.
\item The multilayer structure enhances transient diversity while maintaining convergence.
\end{itemize}

These numerical results support the theoretical findings regarding convergence under time-varying and partially disconnected interaction structures.

\section{Discussion and Further Study}

The proposed multilayer garbage disposal game (GDG) framework provides a unified structure for studying distributed redistribution dynamics over time-varying interaction graphs. The model incorporates two key constructions for interaction matrices: the Metropolis construction and the uniform redistribution rule. Both yield doubly stochastic update matrices, ensuring mass conservation at each time step while permitting heterogeneous local redistribution across network structures.

A central finding of this work is that consensus emerges under intermittent connectivity assumptions. In particular, connectivity of the interaction graphs is not required at every time step; instead, it is sufficient that connected structures occur infinitely often along the switching process. This condition highlights the robustness of the multilayer GDG dynamics with respect to structural uncertainty and temporary disconnections in the underlying network.

From a modeling perspective, the Metropolis construction corresponds to degree-balanced pairwise redistribution, while the uniform redistribution rule distributes mass uniformly over the edge set of the instantaneous graph. Although these constructions differ in local weighting mechanisms, both preserve stochasticity and lead to global averaging behavior under the same connectivity principles. The multilayer extension further introduces cross-layer coupling through averaging operations, which enriches transient dynamics while maintaining asymptotic convergence to a common global average state.

The numerical experiments confirm the theoretical results established in Theorems~\ref{Thm:ConsensusUnderM1}--\ref{Thm:ConsensusUnderM2} and Corollaries~\ref{Cor:ConsensusUnderDistributeThenAverage}--\ref{Cor:ConsensusUnderAverageThenDistribute}. In particular, both single-layer and multilayer GDG dynamics converge to the predicted consensus values, despite frequent occurrences of sparse or disconnected graphs. The simulations also illustrate that the multilayer structure increases transient heterogeneity across layers before convergence is achieved.

Several directions for further study naturally arise. First, it would be of interest to relax the independence assumption in the graph switching process and consider temporally correlated or adversarially generated interaction sequences. Second, relaxing the doubly stochasticity constraint on the directed interaction matrices would broaden the applicability of the model to more general resource exchange systems. Third, the impact of heavy-tailed or non-identically distributed initial states on convergence speed and stability remains an open question. Finally, establishing quantitative convergence rates under minimal connectivity conditions would provide a more refined characterization of multilayer GDG dynamics beyond asymptotic convergence.

\section{Statements and Declarations}
\subsection{Competing Interests}
The author is funded by NSTC grant.

\subsection{Data availability}
No associated data was used.

\begin{thebibliography}{10}

\bibitem{boccaletti2014structure}
Stefano Boccaletti, Ginestra Bianconi, Regino Criado, Charo~I Del~Genio, Jes{\'u}s G{\'o}mez-Gardenes, Miguel Romance, Irene Sendina-Nadal, Zhen Wang, and Massimiliano Zanin.
\newblock The structure and dynamics of multilayer networks.
\newblock {\em Physics reports}, 544(1):1--122, 2014.

\bibitem{kivela2014multilayer}
Mikko Kivel{\"a}, Alex Arenas, Marc Barthelemy, James~P Gleeson, Yamir Moreno, and Mason~A Porter.
\newblock Multilayer networks.
\newblock {\em Journal of complex networks}, 2(3):203--271, 2014.

\bibitem{dickison2016multilayer}
Mark~E Dickison, Matteo Magnani, and Luca Rossi.
\newblock {\em Multilayer social networks}.
\newblock Cambridge University Press, 2016.

\bibitem{degroot1974reaching}
Morris~H DeGroot.
\newblock Reaching a consensus.
\newblock {\em Journal of the American Statistical association}, 69(345):118--121, 1974.

\bibitem{jadbabaie2003coordination}
Ali Jadbabaie, Jie Lin, and A~Stephen Morse.
\newblock Coordination of groups of mobile autonomous agents using nearest neighbor rules.
\newblock {\em IEEE Transactions on automatic control}, 48(6):988--1001, 2003.

\bibitem{olfati2004consensus}
Reza Olfati-Saber and Richard~M Murray.
\newblock Consensus problems in networks of agents with switching topology and time-delays.
\newblock {\em IEEE Transactions on automatic control}, 49(9):1520--1533, 2004.

\bibitem{moreau2005stability}
Luc Moreau.
\newblock Stability of multiagent systems with time-dependent communication links.
\newblock {\em IEEE Transactions on automatic control}, 50(2):169--182, 2005.

\bibitem{olfati2007consensus}
Reza Olfati-Saber, J~Alex Fax, and Richard~M Murray.
\newblock Consensus and cooperation in networked multi-agent systems.
\newblock {\em Proceedings of the IEEE}, 95(1):215--233, 2007.

\bibitem{seneta2006non}
Eugene Seneta.
\newblock {\em Non-negative matrices and Markov chains}.
\newblock Springer Science \& Business Media, 2006.

\bibitem{hajnal1958weak}
John Hajnal and Maurice~S Bartlett.
\newblock Weak ergodicity in non-homogeneous markov chains.
\newblock In {\em Mathematical Proceedings of the Cambridge Philosophical Society}, volume~54, pages 233--246. Cambridge University Press, 1958.

\bibitem{wolfowitz1963products}
Jacob Wolfowitz.
\newblock Products of indecomposable, aperiodic, stochastic matrices.
\newblock {\em Proceedings of the American Mathematical Society}, 14(5):733--737, 1963.

\bibitem{touri2012product}
Behrouz Touri.
\newblock {\em Product of random stochastic matrices and distributed averaging}.
\newblock Springer Science \& Business Media, 2012.

\bibitem{touri2010ergodicity}
Behrouz Touri and Angelia Nedic.
\newblock On ergodicity, infinite flow, and consensus in random models.
\newblock {\em IEEE Transactions on Automatic Control}, 56(7):1593--1605, 2010.

\bibitem{friedkin1990social}
Noah~E Friedkin and Eugene~C Johnsen.
\newblock Social influence and opinions.
\newblock {\em Journal of mathematical sociology}, 15(3-4):193--206, 1990.

\bibitem{castellano2009statistical}
Claudio Castellano, Santo Fortunato, and Vittorio Loreto.
\newblock Statistical physics of social dynamics.
\newblock {\em Reviews of modern physics}, 81(2):591, 2009.

\bibitem{lanchier2022consensus}
Nicolas Lanchier and Hsin-Lun Li.
\newblock Consensus in the {H}egselmann--{K}rause model.
\newblock {\em Journal of Statistical Physics}, 187(3):1--13, 2022.

\bibitem{li2022mHK}
Hsin-Lun Li.
\newblock Mixed {H}egselmann-{K}rause dynamics.
\newblock {\em Discrete and Continuous Dynamical Systems - B}, 27(2):1149--1162, 2022.

\bibitem{li2023mHK2}
Hsin-Lun Li.
\newblock Mixed {H}egselmann-{K}rause dynamics {II}.
\newblock {\em Discrete and Continuous Dynamical Systems - B}, 28(5):2981--2993, 2023.

\bibitem{li2024mixed}
Hsin-Lun Li.
\newblock Mixed {H}egselmann-{K}rause dynamics on infinite graphs.
\newblock {\em Journal of Statistical Mechanics: Theory and Experiment}, 2024(11):113404, 2024.

\bibitem{li2025leader}
Hsin-Lun Li.
\newblock Leader--follower dynamics: Stability and consensus in a socially structured population.
\newblock {\em AIMS Math}, 10(2):3652--3671, 2025.

\bibitem{hirai2006coalition}
Toshiyuki Hirai, Takuya Masuzawa, and Mikio Nakayama.
\newblock Coalition-proof nash equilibria and cores in a strategic pure exchange game of bads.
\newblock {\em Mathematical Social Sciences}, 51(2):162--170, 2006.

\bibitem{li2025multilayer}
Hsin-Lun Li.
\newblock The multilayer garbage disposal game.
\newblock {\em Journal of Difference Equations and Applications}, 31(8):1162--1168, 2025.

\end{thebibliography}

\end{document}